\documentclass[12pt]{iopart}
\usepackage{amsfonts}
\usepackage{epsfig}
\newcommand{\sh}{\mathop{\mathrm{sh}}}
\newcommand{\bee}{\begin{eqnarray}}
\newcommand{\eend}{\end{eqnarray}}
\begin{document}

 \vspace{1cm} \title[Black-hole concept of a supercritical
 nucleus]
 {Black-hole concept of a
point-like nucleus with supercritical charge.}
\author{A E Shabad}
\address{I.E.Tamm Department of Theoretical Physics, P.N.Lebedev Physical
Institute,   Russian Academy of Sciences, Leninsky prospekt 53,
Moscow 117924} \ead{shabad@lpi.ru}
\begin{abstract}
The Dirac equation for an electron in the central Coulomb field of
a point-like nucleus with the charge greater than 137 is
considered. This singular problem, to which the fall-down onto the
centre is inherent, is addressed using a new approach, basing on a
black-hole concept of the singular centre and capable of producing
cut-off-free results. To this end the Dirac equation is presented
as a generalized eigenvalue boundary problem of a self-adjoint
operator. The eigenfunctions make complete sets, orthogonal with a
singular measure, and describe particles, asymptotically free and
 delta-function-normalizable both at infinity and near the singular centre
$r=0$. The barrier transmission coefficient for these particles
responsible for the effects of electron absorption and spontaneous
electron-positron pair production is found analytically as a
function of electron energy and charge of the nucleus. The
singular threshold behaviour of the corresponding amplitudes
substitutes for the resonance behaviour, typical of the
conventional theory, which appeals to a finite-size nucleus.
 %\submitto{\JPA}
\end{abstract}
\section{Introduction}
 The radial Dirac
equation for an electron in the Coulomb potential
\bee\label{coulomb}-\alpha Z/r, \eend of a point-like nucleus,
where $\alpha$ is the fine-structure constant $\alpha=1/137$ and
$Z$ is the nucleus charge (for formal purposes negative values of
$Z$ will be also included into our consideration), is a set of two
first-order differential equations \cite{akhiezer}, \cite{blp}. It
is convenient to write it in the following matrix form
\bee\label{rad} \mathcal{L}\Psi(r)=\left(\varepsilon
+\frac{Z\alpha}{r}\right)\Psi(r),\quad r\in (0,\infty),\eend
\bee\label{L} \mathcal{L}=-\rmi\sigma_2\frac\rmd {\rmd
r}+\frac\kappa{r}\sigma_1+m\sigma_3.\eend Here the wave function
is the two-component
spinor\bee\label{Psi}\Psi(r)=\left(\begin{tabular}{c}$
G(r)$ \\
 $F(r)$
\end{tabular}\right),\eend  $\sigma_i$ are the Pauli matrices:
 \bee\label{pauli} \sigma_1=\left(\begin{tabular}{cc} 0 &1\\1 &
 0\\
\end{tabular}\right),\qquad \rmi\sigma_2=\left(\begin{tabular}{cc} 0 &1\\-1&
0\\
\end{tabular}\right),\qquad \sigma_3=\left(\begin{tabular}{cc} 1& 0\\0&
-1\\\end{tabular}\right),\eend $m$ is the electron mass, and
$\varepsilon$ is its energy. The quantity $\kappa$ is the orbital
momentum\bee\label{kappa}
\kappa=-(l+1)\quad {\rm for}\quad j=l+\frac 1{2},\nonumber\\
\kappa=l\quad {\rm for}\quad j=l-\frac 1{2}.\eend In what follows
we confine ourselves to the lowest orbital state $\kappa=-1$.
Solutions to equation (\ref{rad}) are known \cite{akhiezer},
\cite{blp} in terms of confluent hypergeometric functions.

We are most interested in the supercritical case $\alpha |Z|>1$,
of which the falling to the centre is characteristic. One can
exclude one component from (\ref{rad}) to reduce it to a
second-order differential equation (see $e.g.$ \cite{akhiezer},
\cite{blp}, \cite{zp}, \cite{case}). Then the Coulomb term
(\ref{coulomb}) in (\ref{rad}) gives rise to the
inverse-radius-squared singular term $-(\alpha Z/r)^2$ in the
potential, which is negative (attraction) irrespective of the sign
of $\alpha Z$. This illustrates the presence of the fall-down onto
the centre phenomenon inherent to the Dirac equation (\ref{rad})
with $\alpha|Z|>1$. This case is of direct physical importance as
introducing the effect \cite{kolesnikov}, \cite{gershtein},
\cite{pieper}, \cite{zp}, \cite{greiner} of spontaneous
electron-positron pair creation by an overcharged nucleus created
for a short time in heavy ion collisions (see the review
\cite{pokotilov}), and also the effect of  strong absorption of
electrons by this nucleus - see our previous paper \cite{shabad4}.
(Attribution
 of absorbing quality to a singular centre in other problems and within
other approaches may be found in \cite{alliluev}, \cite{denschlag}.)
Unlike \cite{shabad4}, we find it now more appropriate to be
dealing directly with the set of two first-order differential
equations (\ref{rad}).

To handle the problem, with the supercritical case included, we define
$\mathcal{L}$ (\ref{L}) as a self-adjoint operator, with the Dirac
equation (\ref{rad}) presenting a generalized eigenvalue problem
for it, and identify physical states with its eigen-vectors.
Stress, that $\mathcal{L}$ is not the Hamiltonian, defined as
$\mathcal{H}=\mathcal{L}-\frac{\alpha Z}{r}$. There are many ways
to present a given differential equation as a generalized
eigenvalue problem for different operators. Our choice (\ref{L})
of $\mathcal{L}$ is mainly dictated by the demand that the term,
responsible for the falling to the centre, be kept in the r.-h.
side in eq.(\ref{rad}), to later show itself as a singularity of
the measure, serving the scalar products in the space, spanned by
solutions of the generalized eigenvalue problem. In other words,
our procedure partially extracts the singularity from the
interaction and places it into the measure. The singularity of the
measure offers the physical concept of the singular centre as a
sort of a black hole, emitting and absorbing particles, which are
\textit{free} near the centre and belong to the $continuum$ of
eigenvectors of $\mathcal{L}$.

This is different from what may be obtained using the self-adjoint
extension of the Hamiltonian within the von Neuman technique,
resulting in a discrete spectrum, unlimited from below, for the
Schr$\ddot{\rm o}$dinger equation case \cite{case}, \cite{meetz},
and in a discrete spectrum, condensing near the
infinitely-strong-binding point ~$\varepsilon =-m$, for the Dirac
equation case \cite{case}.

The present work is an extension to the Dirac equation of
analogous treatment, developed earlier \cite{shabad3},
\cite{shabad} for the (second-order) Schr$\ddot{\rm o}$dinger
equation with singular potential with inverse-square singularity.
We find that the extension to the relativistic case is
straightforward, with the only complication due to the spinorial
character of the Dirac equation. This is the lack of positive
definiteness of the measure in the definition of scalar products
for negative energy and positive charge. It is shown, however,
that one can easily restrict oneself to vectors with a definite
sign of the norm within one set, and also exclude zero-norm
vectors by imposing a certain easy-to-observe convention
concerning the way the ends of the interval ~$(r_{\rm L},r_{\rm
U})$, where equation (\ref{rad}) is defined - first taken finite -
tend to their final positions in the singularity points $r=0$ and
$r=\infty$.

In Section \ref{Bliss} we set the generalized self-adjoint
eigenvalue problem associated with the Dirac equation in a
two-side-limited box
 $~~r_{\rm L}<r<r_{\rm U},$~~$r_{\rm L}\rightarrow 0$,~~$r_{\rm U}
 \rightarrow\infty$~ with zero and (anti)periodic boundary conditions,
  derive the orthogonality relations with a singular
 measure and point a family, labelled by the ratio $R=\alpha Z/\varepsilon$,
  of Hilbert spaces, spanned by solutions of the
 generalized eigenvalue problem, including the ones that are
 $\delta$-function-normalizable and correspond
 to particles, free both near the origin $r=0$ and near $r=\infty$.
  In Section \ref{Dilated} we describe the coordinate
 transformation $\xi(r)$ that reduces the generalized eigenvalue problem
 to the standard one by mapping the origin $r=0$ to infinitely remote point
  $\xi=\pm\infty$ (depending on the sign of the energy). In the new coordinate
  $\xi$ the free character of the wave function behavior near the singular
  point $r=0$ becomes explicit, and its normalization to $\delta$-function
  becomes standard. Referring to the one-dimensional barrier problem on the
  infinite axis $\xi$, to which the Dirac
  equation is reduced in our procedure, we find in Section
  \ref{Transmission}
   the reflection and transmission coefficients for the waves, reflected
   from and transmitted to the singularity in $r=0$. It is important, that in
   the Dirac equation, in contrast to the Klein-Gordon case, the function,
   inverse
   to the transformation $\xi (r)$, is
two-valued in the negative-energy continuum. This fact leads to the change
of identification of the incident and reflected waves near $r=0$ as compared
 to the positive-energy domain, and  results in absence of the
 superradiation phenomenon, described in \cite{chandra}, \cite{davies}.
 In Section \ref{Absorption} the transmission coefficient is interpreted
 within the Dirac-sea picture as probability of absorption of electrons by
 the supercritical nucleus in the positive-energy range, and as distribution
  of spontaneously produced electrons - which outgo to the singular point
  $r=0$ - and positrons - which outgo to the singular point $r=\infty$.
   These coefficients are studied in detail and plotted as  functions of
   the nucleus charge and electron energy. Their singular threshold behavior
    near the point $\alpha^2Z^2=1$
   is established. In concluding Section \ref{Discussion}, we comment on a
   comparison with the known results \cite{zp}, \cite{greiner} about the
   spontaneous pair creation,
    obtained using the finite nucleus size cut-off, and discuss the lines,
     along which  the second-quantized theory of unstable vacuum
     \cite{nikishov}, \cite{gitman}, \cite{davies} could be applied
     to the supercritical nucleus, basing on the consideration made
     in the present
     paper.

\section{Bliss eigenvalue problem}\label{Bliss}

 There are two singular points of the set
of  differential  equations (\ref{rad}), $r=\infty$ and $r=0$.

The fundamental solutions behave near the point $r=\infty$
as\bee\label{asympinfty}\rme^{\mp\rmi p r}~r^{\mp\rmi\zeta},\quad
\eend where $p=\sqrt{\varepsilon^2-m^2}$,
$\zeta=\frac{\alpha Z\varepsilon}{p}$.

It will be convenient for our present purposes, to present the
well-known facts about this singular point in the following words
.

When $|\varepsilon|<m$, the growing solution should be
disregarded, whereas the other decreases and is thus localized in
the finite region of the $r$-space, far from the singularity point
$r=\infty$. We refer to this situation  by saying: the singularity
in the infinitely remote point repulses the particle.

When $|\varepsilon|>m$, the both solutions (\ref{asympinfty})
oscillate. They are $\delta$-function-normalizable, because are
concentrated mostly near the infinitely remote singular point. We
refer to this situation by saying, that this singularity attracts
particles.  The particles, captured by the infinitely remote
singularity are asymptotically free in the sense that the
fundamental solutions and the spectrum do not depend on $Z$,
$\kappa$ (if it is not fixed), in other words, on any terms in
(\ref{rad}) that do not survive in the limit $r\rightarrow\infty$,
as compared to the constant term $\varepsilon$, responsible for
the singularity in the infinity.  The solutions with positive and
negative signs in the exponentials in (\ref{asympinfty}), the incoming and
outgoing waves in customary wording, may be thought of as ones,
emitted and absorbed by the infinitely remote singularity.

Now we turn to the other singular point of the set of differential
equations (\ref{rad}), the one in the origin $r=0$, and shall
treat it in exactly the same manner. Two fundamental solutions to
equation (\ref{rad}) behave near the origin $r\rightarrow 0$
like\bee\label{like} r^{\pm\rmi\gamma},\eend where
$\gamma=\sqrt{\alpha^2Z^2-1}$.

When $\alpha |Z|<1$, the growing solution should be disregarded,
whereas the other decreases as $r\rightarrow 0$ and is thus
localized  far from the singularity point. We refer to this
situation  by saying: the singularity in the origin repulses the
particle.

When $\alpha |Z|>1$ (supercritical charge), the both solutions
(\ref{like}) oscillate, which is typical of the problems with
singular attractive potential. Within a generalized eigenvalue
problem for the operator associated with the differential
expression $\mathcal{L}$ (\ref{L}), formulated below, they are
$\delta$-function-normalizable, because the measure characteristic
of this problem appears to be singular in the origin, this fact
providing the necessary divergence of the norm. The particle is
thus localized mostly near the origin, attracted by the
singularity. The particles, attracted by the singular center, are
asymptotically free in its vicinity in the sense that the
fundamental solutions and the spectrum do not depend on
$\varepsilon$, $m$, in other words, on any terms in (\ref{rad})
that do not survive in the limit $r\rightarrow 0$, as compared to
the singular potential. The solutions with positive and negative
signs in the exponentials (\ref{like}),  may be thought of as waves,
 emitted and
absorbed by the singular centre.

With the ratio \bee\label{ratio} R=\frac{\alpha
Z}{\varepsilon}\eend fixed, equation (\ref{rad}) can be
written either as\bee\label{rhs1} \mathcal{L}\Psi(r)=\varepsilon
\left(1+\frac R{r}\right)\Psi(r)\eend or as
\bee\label{rhs2} \mathcal{L}\Psi(r)=\alpha Z\left(\frac 1{R}+\frac
1{r}\right)\Psi(r).\eend Thus, equation (\ref{rad}) becomes
a generalized eigenvalue problem of the type, where the eigenvalue
(this is either $\varepsilon$ for (\ref{rhs1}) or $\alpha Z$ for
(\ref{rhs2})) is multiplied by a function of the variable $r$, and
not by unity. In the case of a second-order equation an analogous
eigenvalue problem was first studied by E. Kamke \cite{kamke}- we
called it Kamke eigenvalue problem when using it as applied to the
singular Schr$\ddot{\rm o}$dinger equation in \cite{shabad},
\cite{shabad4}, - but for the set like (\ref{rad}) it was
considered, according to \cite{kamke}, much earlier by G.A.Bliss
\cite{bliss}.

The equation, in which the sign in front of the derivative term is
reversed and, besides, all the matrices are transposed, is called
an $adjoint$ equation. $Equation~~ (\ref{rad})~~ is~~
self-adjoint$ in the sense that solutions  $\overline{\Psi}(r)$ of
its adjoint equation are linearly expressed in terms of its
solutions with the help of a nondegenerate matrix, namely:
$\overline{\Psi}(r)=\sigma_2\Psi(r), ~~\det\sigma_2\neq 0$. $The~~
~~eigenvalue~~ problem~~ (\ref{rhs1})~~ or~~ (\ref{rhs2})~~ is~~
self-adjoint$, provided appropriate boundary conditions are
imposed. As such, it is sufficient to use the zero and periodic
(or antiperiodic) boundary conditions at the ends of the interval.

We shall  consider equation (\ref{rhs1}), (\ref{rhs2}) in four
intervals: ~$(0,\infty),~(r_{\rm L},\infty),~(0,r_{\rm U}),$\\$
~(r_{\rm L},r_{\rm U})$,~$r_{\rm L},r_{\rm U}>0$~ depending on the
regions - called $sectors$ - within which the parameters
$\varepsilon$ and $\alpha Z$ may lie, and pass to the limit
~~$r_{\rm L}\rightarrow 0,~~r_{\rm U}\rightarrow\infty$~~
afterwards. So, in the end, all the intervals are one interval.
The boundary conditions are:
\bee\label{bound}G(0)=G(\infty)=0,\quad{\rm when }\quad|\alpha
Z|<1,\; |\varepsilon|<m;\quad {\bf Sector~ I} \nonumber\\{\rm the~
interval~ is}\quad 0\leq r\leq\infty\eend \bee\label{boundzero2}
G(0)=G(r_{\rm U})=0,\quad{\rm when }\quad |\alpha Z|<1,\;
|\varepsilon|>m;\quad {\bf Sector~ II}\nonumber\\{\rm the~
interval ~is}\quad 0\leq r\leq r_{\rm U} \eend
\bee\label{boundzero}G(r_{\rm L})=G(\infty)=0,\quad{\rm when
}\quad|\alpha Z|>1,\; |\varepsilon|<m;\quad {\bf Sector~
III}\nonumber\\{\rm the~ interval~ is}\quad r_{\rm L}\leq
r\leq\infty \eend\bee\label{period} G(r_{\rm L})=\pm G(r_{\rm
U}),\; F(r_{\rm L})=\pm F(r_{\rm U}),\quad{\rm when }\quad |\alpha
Z|>1,\; |\varepsilon|>m;\nonumber\\ {\rm the ~interval ~is}\quad
r_{\rm L}\leq r\leq r_{\rm U}\qquad \qquad \qquad \qquad \quad
{\bf Sector~ IY}. \eend
 As long as $r_{\rm L}\neq 0$ for $\alpha |Z|>1$ and/or
 $r_{\rm U}\neq\infty$ for
$|\varepsilon|>m$, there are infinite countable manifolds of
eigenvalues in sectors II, III, IV, which condense in the limit
~~$r_{\rm L}\rightarrow
0,~~r_{\rm U}\rightarrow\infty$ to make continua.

The function
$~~\varepsilon + \frac{\alpha Z}{r}~~$  does not change sign
throughout the interval $(0,\infty)$, when $R>0$, $i.e.$ in the
positive energy
domain ($\varepsilon >0$) for positive charge $Z>0$, and in the
negative energy domain ($\varepsilon <0$) for negative charge
$Z<0$. In these regions $\textit{the self-adjoint boundary problem
is positively definite}$. In this case the eigenvalues are real
and the corresponding solutions make, for each value of $R$, a
complete set of states, mutually orthogonal with the measure
\bee\label{measureIV}\rmd\mu(r)=\left(1+\frac R{r}\right)\rmd r
.\eend Below we shall consider also negative $R$ and see, what
 happens in that case. Note, that the e$^+$e$^-$- pairs production
 just occurs in the region $R,0$ of sector IY.

The domains, pointed in (\ref{bound}), (\ref{boundzero2}),
(\ref{boundzero}), (\ref{period}), correspond to what was called
sectors I, II, III, IV in \cite{shabad}, \cite{shabad3}. Sector II
($\alpha |Z|<1,~|\varepsilon|>m$) corresponds to particles, free
at infinity in the limit $r_{\rm U}\rightarrow\infty$ and repulsed
from the centre. Only one of two fundamental solutions belongs to
$L^2_\mu(0,r_{\rm U})$, the space of functions square-integrable
with the measure (\ref{measureIV}) in the interval $(0,r_{\rm
U})$. Sector III ($\alpha |Z|>1,~|\varepsilon|<m$) contains
particles, free in the origin in the limit $r_L\rightarrow 0$ and
repulsed from the infinitely remote point. Only one fundamental
solution belongs to $L^2_\mu(r_{\rm L},\infty))$. In sector IV
($\alpha |Z|>1,~|\varepsilon|>m$) particles are free both near the
origin and near the infinity in the limit $r_{\rm L}\rightarrow
0$, $r_{\rm U}\rightarrow \infty$. The both fundamental solutions
are $L^2_\mu(r_{\rm L},r_{\rm U})$. The (anti)periodic boundary
conditions (\ref{period}) are imposed in agreement with the
inelastic character of the scattering in sector IV, where there is
a nonzero current inflow through the outer border $r=r_{\rm U}$,
equal to the outflow through the inner border $r=r_{\rm L}$, and
$vice~versa$. As distinct from this, in sectors II and III the
current is zero, the scattering of particles emitted by the
infinity (sector II) and by the centre (sector III) is elastic:
everything what is emitted by the centre (infinity) is reflected
back to the centre (infinity). As for sector I ($\alpha
|Z|<1,~|\varepsilon|<m$), where the infinite countable manifold of
hydrogen-like bound states lies in
the quadrants  ($0<\alpha Z<1,~0<\varepsilon<m$) and ($-1<\alpha
Z<0,~-m<\varepsilon<0$), neither fundamental solution generally
belongs to
$L^2_\mu(0,\infty)$.

To be more explicit, let us - following the standard way -
left-multiply the equation
(\ref{rhs1}) (or (\ref{rhs2})) by the spinor $\Psi^*_1(r)$, which
is the solution of the complex-conjugate equation, but with
$\varepsilon_1$ taken instead of $\varepsilon$ (or $\alpha Z_1$
instead of $\alpha Z$) and the same $R$. (Stress that all the
coefficients and matrices in expression (\ref{L}) are real and do
not depend either on $\varepsilon$ or on $\alpha Z$). Let us,
next, subtract the same product, with $\Psi(r)$ and $\Psi^*_1(r)$
interchanged and integrate the difference over $r$ within the
limits $r_1<r<r_2$. In agreement with (\ref{bound}),
(\ref{boundzero2}), (\ref{boundzero}), (\ref{period}), in sectors I
and II  the lower limit  should be chosen as $r_1=0$, whereas in
sectors III and IV it is $r_1=r_L$. The upper limit $r_2$
coincides with $r_U$ in sectors II and IV, and is infinite
$r_2=\infty$ in sectors I and III. Then one has \bee\label{green}
-\rmi\left.\Psi^*_1(r)\sigma_2\Psi(r)\right|^{r_2}_{r_1}=
(\varepsilon_1^*-\varepsilon)
\int^{r_2}_{r_1}\Psi^*_1(r)\Psi(r)\left(1+\frac R{r}\right)\rmd
r=\nonumber\\
=\alpha(Z_1^*-Z) \int^{r_2}_{r_1}\Psi^*_1(r)\Psi(r)\left(\frac
1{R}+\frac 1{r}\right)\rmd r.\eend When written explicitly
in components this equation looks like:\bee\label{greenC}
\left.(F^*_1(r)G(r)-F(r)G^*_1(r))\right|^{r_2}_{r_1}=\nonumber\\
=(\varepsilon_1^*-\varepsilon)
\int^{r_2}_{r_1}(G(r)G^*_1(r)+F(r)F^*_1(r))\left(1+\frac
R{r}\right)\rmd
r=\nonumber\\
=\alpha(Z_1^*-Z)
\int^{r_2}_{r_1}(G(r)G^*_1(r)+F(r)F^*_1(r))\left(\frac 1{R}+\frac
1{r}\right)\rmd r.\eend

The boundary conditions (\ref{bound}), (\ref{boundzero}),
(\ref{boundzero2}), (\ref{period}) provide the vanishing of the
left-hand side of (\ref{green}) or  (\ref{greenC}) and make the
differential expression $\mathcal{L}$ a self-adjoint operator. By
omitting the index 1 in this relation we obtain that for the
eigenvalues $\varepsilon$ and $Z$ of the self-adjoint operator
$\mathcal{L}$ to be real \bee\label{real} \varepsilon =
\varepsilon^*,\quad Z=Z^* \eend it is sufficient that the function
~~$(\frac 1{R}+\frac 1{r})$~~ should have a definite sign, since
in this case the integral in (\ref{green}) is nonzero. Such
situation occurs when $R>0$.

Otherwise, when $R<0$, the zero-norm vectors, satisfying the
relation \bee\label{zero}
\int^{r_2}_{r_1}\Psi^*(r)\Psi(r)\left(1+\frac R{r}\right)\rmd r=0
\eend might exist and then the eigenvalues corresponding to
it be not necessarily real. Also negative-norm solutions might
appear with real eigenvalues, and the manifold of eigenfunctions
of the self-adjoint operator (\ref{L}) might not be complete,
according to \cite{kamke}, \cite{bliss}. In such case it had to be
 completed to the Hilbert space, as explained in \cite{nagy}. As
a matter of fact, this scenario can be avoided.

 It follows from (\ref{greenC}), (\ref{bound}), (\ref{boundzero2}),
(\ref{boundzero}),
(\ref{period}),  that the eigenfunctions with nonzero
norm obey  the orthogonality relations
\bee\label{orth}\int^{r_2}_{r_1}\Psi^*_k(r)\Psi_n(r)\left(1+\frac
R{r}\right)\rmd r=N\delta_{kn}.\eend Here the integers
$k,n$ label the eigenfunctions belonging to different countable
eigenvalues $\varepsilon$ or $Z$, and the wave functions $\Psi$
relate to a fixed value of $R$. The norm $N$ in every sector,
except sector I, gets
predominant contribution from the integration near the points
$r=r_{\rm L}$ and $r=r_{\rm U}$.

Bearing in mind the asymptotic behaviours (\ref{asympinfty}) and
(\ref{like}), and the boundary conditions (\ref{period}), we get
for the eigenfunctions in sector IV in the asymptotic regime of
very large upper size of the interval $r_{\rm U}\gg|R|, m^{-1}$
and very small lower size $r_{\rm L}\ll|R|, m^{-1}$ - up to a
finite factor - that\bee\label{norm}N^{IV}\asymp\int^{r_{\rm
U}}_{r_{\rm L}}\left(1+\frac R{r}\right)\rmd r\asymp r_{\rm
U}+R\ln\frac{r_0}{r_{\rm L}}. \eend  Here $r_0$ is an arbitrary
positive dimensional constant, $r_{\rm L}\ll r_0\ll r_{\rm U}$. We
neglected $r_{\rm L}$ as compared to ~$|R\ln\frac{r_0}{r_{\rm
L}}|$,~ and ~$|R\ln\frac {r_{\rm U}}{r_0}|$~ as compared to
$r_{\rm U}$. A change of $r_0$ does not violate eq. (\ref{norm})
within its accuracy.

For positive $R$ the norm (\ref{norm}) is positive. For negative
$R$ the norm (\ref{norm}) \textit {has a definite sign} for a
definite value of $R$, provided either of the inequalities
\bee\label{inequality}\frac{r_{\rm U}}{|R|}>\ln\frac{r_0}{r_{\rm
L}} \quad {\rm or}\quad\frac{r_{\rm U}}{|R|}<\ln\frac{r_0}{r_{\rm
L}}\eend is observed during the limiting process $r_{\rm
L}\rightarrow 0,~~r_{\rm U}\rightarrow\infty$. By imposing this
additional requirement we avoid the appearance of vectors with
different signs of the norm within the same set of eigenfunctions.
The zero-norm vectors of state do not appear either. Thus, the
problem of indefinite metric is eliminated in sector IV. In other
sectors it does not appear.

In sector II, where the eigenfunctions are $L_\mu^2(0,r_{\rm U})$,
the integration in (\ref{orth}) near $r=0$ is convergent, while
the upper limit gives a predominant contribution to (\ref{orth})
for  $r_U\gg|R|, m^{-1}$ due to the asymptotic behaviour
(\ref{asympinfty})\bee\label{normII}N^{II}\asymp\int^{r_{\rm
U}}\left(1+\frac R{r}\right)\rmd r\asymp r_{\rm U}. \eend This has
a definite sign. In sector III, where the eigenfunction is
$L_\mu^2(r_{\rm L},\infty)$, the integration in (\ref{orth}) near
the upper limit converges, but near the lower limit gives a
predominant contribution for  $r_{\rm L}\ll|R|,m^{-1}$ due to the
asymptotic behaviour (\ref{like})
\bee\label{normIII}N^{III}\asymp\int_{r_{\rm L}}\left(1+\frac
R{r}\right)\rmd r\asymp R\ln\frac{r_0}{r_{\rm L}}. \eend This has
a definite sign for a given sign of $R$. As for sector I, the
eigenvalue problem (\ref{rhs1}), (\ref{rhs2}) with the boundary
conditions (\ref{bound}) does not have any solutions for $R<0$:
the hydrogen-like bound states only lie in the segments with $R>0$
in this sector. In the latter region, for the known
 discrete values of the energy $\varepsilon$, depending on the charge
 $\alpha Z$, which provide the fulfillment of the boundary condition
 (\ref{bound}), the solution decreases as
 ~exp$\{-r\sqrt{m^2-\varepsilon^2}\}$~ at~$r\rightarrow\infty$
 and as $r^{\sqrt{1-\alpha^2Z^2}}$ at
 $r\rightarrow 0$.
 Correspondingly, the norm
\bee\label{normI}
N^I\asymp\int_0^\infty\Psi^*(r)\Psi(r)\left(1+\frac
R{r}\right)\rmd r<\infty\eend converges in spite of the
singularity in the measure.

It may be seen from the results of \cite{zp}, where the
Klein-Gordon
 equation was considered, that the corresponding measure would
 in that case  be $\rmd\mu(r)=(\frac
1{R}+\frac 1{r})^2\rmd r$, which is positively definite, and hence
the problem of indefinite metrics does not arise.

Manifolds of eigenvalues of the problem (\ref{rhs1}) (or
(\ref{rhs2})) may belong simultaneously to two or three sectors:
to sectors I, II, IV, if $|R|<m^{-1}$, to sectors I, III, IV, if
$|R|>m^{-1}$, and to sectors I, IV, if $|R|=m^{-1}$.
Correspondingly, the resolution of the unity includes integration
over continua of eigenvectors in all the sectors II, III, IV
crossed by the straight line $R=const$ in the plane $(\varepsilon,
\alpha Z)$, the states of both signs of the charge and energy
being simultaneously involved, as well as summation over bound
states in sector I for $R>0$. To conform the boundary conditions
(\ref{boundzero2}) in sector II and (\ref{period}) in sector IV,
what is needed when $0>R>-m^{-1}$, we have to choose the left
inequality in (\ref{inequality}), so that the signs of the norms
be the same - positive - within one set of eigenvalues.
Analogously, when $R<-m^{-1}$, we have to choose the right
inequality in it, so that the boundary conditions
(\ref{boundzero}) in sector III and (\ref{period}) in sector IV
might be in mutual agreement and the sign of the norm be common -
negative in this case. At last, in the special case $R=-m^{-1}$
the problem is indifferent to the choice of the
inequality sign in (\ref{inequality}). %Transitions along the line
%$|R|>m^{-1}$ do not cause the restructuring of the ground state,
%since we do not encounter different densities of continua when
%crossing the border between different sectors. One may think
% that the transition is adiabatic in this case, not escaping the
% thermal equilibrium and corresponds to maximum entropy.
 The unrestricted growth of the norm (\ref{norm}) in the limit
$r_{\rm L}\rightarrow 0,~~r_{\rm U}\rightarrow\infty$ gives the
possibility to replace the r.-h. side of (\ref{orth}) by
$\delta$-function of $~~(\varepsilon_1-\varepsilon)~~$ or of
$~~(Z_1-Z)~~$.

The integrals (\ref{norm}), (\ref{normII}), (\ref{normIII}) play
the role of effective sizes of the quantization boxes in the
corresponding sectors. Once the spacings between the levels prior
to the limiting transition $r_{\rm L}\rightarrow 0$ and/or $r_{\rm
U}\rightarrow \infty$ are inversely proportional to these
effective sizes ($cf$ \cite{shabad3}), we conclude that the
densities of states are essentially different in different
sectors, which is important for physical processes occurring when
a change of parameters  causes transitions between sectors, as in
heavy ion collisions.

Besides the family of Hilbert spaces labelled by the ratio $R$
(\ref{ratio}), we can point two more families. The standard one is
associated with the representation of the Dirac equation
(\ref{rad}) in the form of the eigenvalue problem for the
Hamiltonian\bee\label{hamiltonian}
\mathcal{H}=\mathcal{L}-\frac{\alpha Z}{r},\eend the operator,
which leaves aside the term $\varepsilon\Psi$ of the Dirac
equation, singular near $r=\infty$, but includes all the terms
singular in the origin \bee\label{radH} \mathcal{H}
\Psi(r)=\varepsilon\Psi(r).\eend This eigenvalue problem is
well-defined within our procedure, which is standard in this case,
when there is no singular attractiveness in the origin, $i.e.$ for
$\alpha |Z|<1$. The families of eigen-vectors, orthogonal for
different values of $\varepsilon$ with the measure $\rmd r$, are
labelled by the charge $-137<Z<137$ and consist for each charge of
the wave functions corresponding to discrete bound states in
sector I, $-m<\varepsilon<m$, and two continua of positive,
$\varepsilon>m$, and negative, $\varepsilon<-m$, energies in
sector II.

 The
other one is associated with the representation of the Dirac
equation (\ref{rad}) in the form of the generalized eigenvalue
problem for the operator $\mathcal{L}-\varepsilon$, which leaves
aside the term $\frac{\alpha Z}{r}\Psi$ of the Dirac equation,
singular near $r=0$, but includes all the terms singular in the
infinitely remote point,
\bee\label{radLIII}(\mathcal{L}-\varepsilon) \Psi(r)=\frac{\alpha
Z}{r} \Psi(r).\eend This eigenvalue problem is well-defined within
our procedure, when there is no singular attractiveness in
infinitely remote point, $i.e.$ for $|\varepsilon |<m$. The
families of eigen-vectors, orthogonal for different values of $Z$
with the measure $\frac{\alpha }{r}\rmd r$, are labelled by the
energy $-m<\varepsilon<m$ and consist for each energy of the wave
functions corresponding to discrete bound states in sector I,
$-137<Z<137$, and two continua of positive, $Z>137$, and negative,
$Z<-137$, charges in sector III.
\section{Dilated coordinate space}\label{Dilated}
The generalized eigenvalue problem (\ref{rhs1}) (or (\ref{rhs2}))
can be formally reduced to the form of the standard eigenvalue
problem like (\ref{radH})
\bee\label{standard}-\rmi\sigma_2\frac{\rmd\widetilde
{\Psi}(\xi)}{\rmd\xi}+\frac {r
(\xi)}{r(\xi)+R}\left(\frac{-\sigma_1}{r
(\xi)}+m\sigma_3\right)\widetilde{\Psi}(\xi)
=\varepsilon\widetilde{\Psi}(\xi) \eend
 with the
help of the dilatation of the variable\bee\label{ksi}
\xi(r)=\int_{r_0}^r\left(1+\frac{R}{r}\right) \rmd r= R
\ln\frac r{r_0}+r-r_0.\eend The positive
dimensional constant $r_0$ here is arbitrary.

If $R>0$, the transformation (\ref{ksi}) is a monotonous function
of $r$, and the function $r(\xi)$, inverse to $\xi(r)$, is
single-valued. In this case the spinor in (\ref{standard}) should
be understood as $\widetilde{\Psi}(\xi)=\Psi(r(\xi))$, and the
interval, where this equation is defined, is
$\xi\in(-\infty,\infty)$. The denominator in the second term of
(\ref{standard}) is nonzero.

If, however, $R<0$, the denominator turns to zero in the point
$r=-R$, and also the function $r(\xi)$,  inverse to $\xi(r)$, is
two-valued. Denote the two branches of the inverse function as
$r_{\rm a}(\xi)$ and $r_{\rm b}(\xi)$:\bee\label{branches}0<r_{\rm
a}(\xi)<-R,\quad \xi_{\rm min}<\xi<\infty\nonumber\\-R<r_{\rm
b}(\xi)<\infty,\quad \xi_{\rm min}<\xi<\infty.\eend Due to
(\ref{ksi}) the extremum condition $(\rmd\xi/\rmd r)=0$ is
satisfied in the point $r=-R$, hence $\xi_{\rm min}=\xi(-R)$. The
value of $\xi_{\rm min}$ contains arbitrariness due to the
arbitrariness of $r_0$. Its specific choice is not important. Now
(\ref{standard}) becomes two
equations\bee\label{eqAB}-\rmi\sigma_2\frac{\rmd\widetilde
{\Psi}_{\rm a,b}(\xi)}{\rmd\xi}+\frac {r_{\rm a,b} (\xi)}{r_{\rm
a,b}(\xi)+R}\left(\frac{-\sigma_1}{r_{\rm a,b}
(\xi)}+m\sigma_3\right)\widetilde{\Psi}_{\rm a,b}(\xi)
=\varepsilon\widetilde{\Psi}_{\rm a,b}(\xi),\nonumber\\
\xi_{\rm min}\leq\xi\leq\infty\eend for the two spinors
$\widetilde{\Psi}_{\rm a,b}(\xi)=\Psi(r_{\rm a,b}(\xi))$, defined
in the same interval of $\xi$, but with two different functions
$r_{\rm a}(\xi)$ and $r_{\rm b}(\xi)$ that coincide in the end-point
$r_{\rm a,b}(\xi_{\rm min})=-R$. Each of these equations has
a singularity at the end of the interval $\xi=\xi_{\rm min}$. The
two spinors are connected by the continuity
relation\bee\label{branch} \widetilde{\Psi}_{\rm
a}(-R)=\widetilde{\Psi}_{\rm b}(-R). \eend

The current
\bee\label{current}j(r)=\rmi\Psi^*(r)\sigma_2\Psi(r)=F(r)G^*(r)
-F^*(r)G(r)\eend is conserved, $(\rmd j/\rmd r)=0$, on the interval
$0<r<\infty$ due to eq.(\ref{rhs1}). Under the transformation
(\ref{ksi}) it turns into two currents $\widetilde{j}_{\rm a,b}
(\xi)=j(r_{\rm a,b}(\xi))=\rmi\widetilde{\Psi}_{\rm a,b}
^*(\xi)\sigma_2\widetilde{\Psi}_{\rm a,b}(\xi)$ that both
conserve, $(\rmd \widetilde{j}_{\rm a,b}/\rmd\xi)=0$,  as a
consequence of eqs.(\ref{eqAB}), in the interval $\xi_{\rm
min}<\xi<\infty$. Eq.(\ref{branch}) guarantees the coincidence of
the two currents in the singular end-point $\xi=\xi_{\rm min}$. We
conclude that the current does not undergo a discontinuity when
passing the singularity point in equations (\ref{eqAB}), once the
latter are obtained from (\ref{rhs1}).

The transformation (\ref{ksi}) maps the point $r=\infty$ to
$\xi=\infty$, and the point $r=0$ to either positive or negative
infinity, depending on the sign of the ratio $R$ (\ref{ratio}). In
terms of the delated variable $\xi$ the asymptotic behaviour
(\ref{like}) - up to the coordinate-independent factor
$r_0^{\pm\rmi\gamma}$ -
becomes\bee\label{likexi}\exp{(\pm\rmi\xi\frac\gamma{R})},\quad
|\xi|\rightarrow\infty. \eend This looks like an ordinary free
wave in the $\xi$-space with the $pseudo$momentum
$(\gamma/R)=\pm\sqrt{\varepsilon^2-R^{-2}}$. (For the special
value $R=-m^{-1}$, noted above, the psedomomentum coincides with
the momentum.) Under the transformation (\ref{ksi}) the measure
(\ref{measureIV}) turns into $\rmd\mu(r)=\rmd\xi$, and the
orthogonality relation (\ref{orth}) now
is\bee\label{orthxi}\int^{\xi(r_2)}_{\xi(r_1)}\widetilde{\Psi}^
*_k(\xi)\widetilde{ \Psi}_n(\xi)d\xi=N\delta_{kn}.\eend The norm
(\ref{norm}) in sector IV becomes \bee N^{IV}=\int^{\xi_{\rm
U}}_{-\xi_{\rm L}}\rmd\xi=\xi_{\rm L}+\xi_{\rm U},\quad {\rm
if}\quad R>0\nonumber\eend
and\bee\label{normxi}N^{IV}\asymp\int_{r_{\rm
L}}^{-R}\left(1+\frac R{r_{\rm a}(\xi)}\right)\rmd r_{\rm
a}(\xi)+\int^{r_{\rm U}}_{-R}\left(1+\frac R{r_{\rm
b}(\xi)}\right)\rmd r_{\rm b}(\xi)=\nonumber\\=\int_{-\xi_{\rm
L}}^{\xi_{\min}}\rmd\xi+\int^{\xi_{\rm
U}}_{\xi_{\min}}\rmd\xi\asymp\xi_{\rm L}+\xi_{\rm U}, \quad {\rm
if}\quad R<0\eend where ~~$\xi_{\rm U}\equiv\xi(r_{\rm U})\asymp
r_{\rm U}\rightarrow\infty$ ~~and ~~$\xi_{\rm L}\equiv -
\xi(r_{\rm L})\asymp R\ln (r_0/r_{\rm L})\rightarrow {\rm
sgn}R\cdot\infty$. Thus, in the representation of the dilated
variable $\xi$ it is explicit that eigenfunctions in sector IV are
normalizable to $\delta$-function, when $\xi_{\rm L}+\xi_{\rm U}$
tends to $positive$ infinity. This occurs, when $R>0$ and when
$-m^{-1}<R<0$. In the latter case $\xi_{\rm L}<0$, but $\xi_{\rm
L}+\xi_{\rm U}\rightarrow +\infty$, once the sign $>$ is chosen in
(\ref{inequality}), as explained in Section 2. On the contrary,
for $R<-m^{-1}$, we agreed above to choose the $<$ sign in
(\ref{inequality}), so that $\xi_{\rm L}+\xi_{\rm U}\rightarrow
-\infty$, and solutions are normalized to $-\delta(\varepsilon
-\varepsilon_1)$. Then the total probabilities should be defined
as norms taken with the minus sign. If, at last, $R=-m^{-1}$,
either convention about the sign in (\ref{inequality}) and,
correspondingly, about the total probability does fit.

Analogously, the norm (\ref{normIII}) in sector III
is\bee\label{normxiIII}N^{III}\asymp\int_{-\xi_{\rm L}}\rmd
\xi=\xi_{\rm L}\rightarrow{\rm sgn} R\cdot\infty.
\eend This is  normalizable to $\delta$-function for $R$
positive, and to minus $\delta$-function for $R$ negative.

The contents of this section perfects the demonstration of
complete symmetry between the inner and outer coordinate spaces,
$r\rightarrow 0$ ($|\xi|\rightarrow\infty$) and
$r\rightarrow\infty $ ($\xi\rightarrow\infty$).

\section{Transmission and reflection coefficients}\label{Transmission}
 In sector IV, free particles emitted by infinitely remote
point (waves, incoming from $r=\infty$) are partially reflected
  backwards, but partially penetrate to the
vicinity of the origin to become free near it (waves, outgoing to
$r=0$), in other words - absorbed by the centre. Also the inverse
process may proceed: free particles emitted by the centre (waves,
incoming from the origin $r=0$) are partially reflected back to
the centre, but partially escape to infinity (waves, outgoing to
$r=\infty$ and "absorbed by the remote point"). Thus, the singular
differential equation is no longer a one-particle problem, but
corresponds to an inelastic, two-channel problem, characterized by
a $2\times2$ scattering matrix, the same as in (\cite{shabad}),
(\cite{shabad3}). Below we shall find the off-diagonal $S$-matrix
element squared - the transmission coefficient that is responsible
for absorption of electrons by the supercritical nucleus, provided
that the electron energy is above $m$, and for spontaneous
elecrton-positron pair production, provided that the energy is
below $-m$. The latter conclusion holds true within the concept of
the Dirac sea - usually appealed to, when this effect is
considered in the customary context (see the review \cite{zp}, and
the monographs \cite{greiner}). The negative-energy free electron,
which at $r=\infty$ belongs to the filled Dirac sea, after it is
scattered off the point-like nucleus, partially transmits to the
centre to become free near it and leaves an empty vacancy in its
stead, interpreted as a positron, free at $r=\infty$. Unlike this
scenario, the one  of refs. \cite{zp}, \cite{greiner} specifies
that the electron near the nucleus is not free, but occupies the
deepest bound state.

In the present section we deal only with positive nucleus charge
$Z$ and are in sector IV \bee\label{sectorIV} \alpha
Z>1,\;\quad|\varepsilon|>m.\eend Consider the fundamental solution
to equation (\ref{rad}), which behaves as a single exponent near
$r=0$\bee\label{left} \left.\Psi(r)\right|_{r\rightarrow 0}\simeq
(2pr)^{\rmi\gamma}\left(\begin{tabular}{c}
{\large$\left(\frac{\rmi (\zeta+\gamma)}{1-\frac{\rmi
m\zeta}{\varepsilon}}+1\right)(\frac
\varepsilon{m}+1)^\frac 1{2}$}\\\\
{\large$ -\left(\frac{\zeta+\gamma}{1-\frac{\rmi
m\zeta}{\varepsilon}}+\rmi\right)(\frac\varepsilon{m}-1)^\frac 1{2}$}\\
\end{tabular}\right),\eend where
\bee\label{parameters}p=\sqrt{\varepsilon^2-m^2},~~\gamma=\sqrt{\alpha^2Z^2-1},~~\zeta=(\alpha
Z\varepsilon)/p,\quad |\zeta|>\gamma\eend are real quantities.
Besides, $|\zeta|>\gamma$. The same solution behaves at $r=\infty$
as\bee\label{ri} \left.\Psi(r)\right|_{r\rightarrow\infty}\simeq
{\rm e}^{-\rmi
pr}(2pr)^{-\rmi\zeta}\left(\begin{tabular}{c}$(\frac\varepsilon{m}+1)^\frac
1{2}$\\$ -\rmi(\frac\varepsilon{m}-1)^\frac
1{2}$\\\end{tabular}\right){\cal {E}} +\nonumber\\+ {\rm e}^{\rmi
pr}(2pr)^{\rmi\zeta}\left(\begin{tabular}{c}$(\frac\varepsilon{m}+1)^\frac
1{2}$\\$ \rmi(\frac\varepsilon{m}-1)^\frac
1{2}$\\\end{tabular}\right){\cal {G}}.\eend The
coordinate-independent coefficients $\mathcal{E}$ and $\mathcal G$
can be determined using the analytical continuation
$\sqrt{1-\alpha^2Z^2}=\rmi\sqrt{\alpha^2Z^2-1}$ of the known exact
solution (see \cite{akhiezer}, \cite{blp}) into supercritical
region $\alpha^2Z^2>1$. These are \bee\label{cal}{\cal
{E}}=\frac{\Gamma (2\rmi\gamma+1)\rme^{-\frac\pi
{2}(\zeta+\gamma)}}{\rmi (\gamma-\zeta)\Gamma
(\rmi(\gamma-\zeta))},\quad {\cal G}=\frac{\Gamma
(2\rmi\gamma+1)\rme^{-\frac\pi {2}(\zeta-\gamma)}}{(1-\rmi\frac
m{\varepsilon}\zeta)\Gamma (\rmi(\gamma+\zeta))}.\eend Calculating
the values of the conserved current (\ref{current}) near $r=0$ and
near $r=\infty$, and equalizing the two values with one another we
obtain the unitarity relation in the form of the
identity\bee\label{unitarity} |{\cal G}|^2=|{\cal
E}|^2+\frac{2\gamma}{\zeta -\gamma}, \eend easy to verify
explicitly by the substitution of (44).

In the positive energy domain, $\varepsilon>m$ ($R>0$),
 in accord with (\ref{likexi}), the exponential in
(\ref{left}) $r^{\rmi\gamma}\simeq\exp{(\rmi\xi\frac\gamma{R})}$
oscillates with the same sign of frequency as the second term in
(\ref{ri}). Referring to the second term in (\ref{ri}) as to the
wave, incoming from infinity, and to (\ref{left}) as the wave
transmitted to the centre, we have to normilize the incoming wave
to unity. This reduces to division of eq.(\ref{unitarity}) over
$|{\cal G}|^2$ to give it the form\bee\label{unity}
1=\mathbb{R_+}+\mathbb{T_+},\eend with $\mathbb{R_+}$ and
$\mathbb{T_+}$ being the reflection and transmission coefficients,
respectively: \bee\label{refltr} \mathbb{R_+}=\frac{|{\cal
E}|^2}{|{\cal G}|^2},\quad \mathbb{T_+}=\frac{2\gamma}{(\zeta
-\gamma )|{\cal G}|^2},\quad \varepsilon>m.\eend As
$\zeta>\gamma$, when $\varepsilon>m$ , one has $\mathbb{T_+}>0$,
hence $\mathbb{R_+}<1$, $\mathbb{T_+}<1$. Finally the following
analytical expression can be deduced for the transmission
coefficient (\ref{refltr}) from (\ref{cal})
\bee\label{transpositive}\mathbb{T_+}=1-\frac{|{\cal E}|^2}{{|\cal
G}|^2}=1-\rme^{-2\pi\gamma}\frac{\sinh \pi(\zeta-\gamma)}{\sinh
\pi(\zeta+\gamma)}=\frac{2\rme^{2\pi\zeta}\sh
2\pi\gamma}{\rme^{2\pi(\zeta+\gamma)}-1}, \qquad
\varepsilon>m.\eend For sufficiently large $\gamma$ this becomes
\bee \mathbb{T_+}=\frac 1{1-\rme^{-2\pi(\gamma
+\zeta)}}.\nonumber\eend Owing to the two-fold character of the
transformation inverse to (\ref{ksi}) in the negative energy
domain, the identification of the incoming and reflected waves is
there different. Now that $\varepsilon<-m$ ($R<0$),
 in accord with (\ref{likexi}), the exponential in (\ref{left})
$r^{\rmi\gamma}\simeq\exp{(\rmi\xi\frac\gamma{R})}$ oscillates
with the same sign of frequency as the $first$ term in (\ref{ri}),
it is the $first$ term that should be referred to as the wave,
incoming from infinity, and the second one as the reflected wave.
Now the normalization of the incoming wave to unity implies the
division of eq.(\ref{unitarity}) over $|{\cal E}|^2$. Then
eq.(\ref{unitarity}) acquires again the  form
(\ref{unity})\bee\label{unity1} 1=\mathbb{R_-}+\mathbb{T_-}, \eend
but this time the reflection and transmission coefficients
are\bee\label {refltrneg} \mathbb{R_-}=\frac{|{\cal G}|^2}{|{\cal
E}|^2},\quad \mathbb{T_-}=\frac{2\gamma}{( \gamma -\zeta)|{\cal
E}|^2},\qquad \varepsilon<-m.\eend In this domain $\zeta<-\gamma$,
and one has $\mathbb{T}_->0$, hence $\mathbb{R_-}<1$,
$\mathbb{T_-}<1$. Finally the following analytical expression can
be found for the transmission coefficient in the negative energy
domain \bee\label{transnegative}\mathbb{T_-}=1-\frac{|{\cal
G}|^2}{{|\cal
E}|^2}=\frac{\mathbb{T_+}}{\mathbb{T_+}-1}=1-\rme^{2\pi\gamma}\frac{\sinh
\pi(\zeta+\gamma)}{\sinh
\pi(\zeta-\gamma)}\nonumber\\
=\frac{1-\rme^{-4\pi\gamma}}{\rme^{-2\pi(\zeta+\gamma)}
-\rme^{-4\pi\gamma}},\qquad \varepsilon<-m.
\eend For sufficiently large $\gamma$ this is
simplified
to\bee\label{simple}\mathbb{T_-}=\rme^{2\pi(\zeta+\gamma)}.\eend The latter expression is valid already for $\alpha
Z-1\gg 1/32\pi^2$.

In our preliminary publication \cite{shabad4} we did not take into
account the fact that the two-valuedness of the coordinate
transformation, inverse to (\ref{ksi}), affects the identification
of incident and reflected waves, and took (\ref{refltr}) as
universal expressions for the reflection and transmission
coefficients valid both in the positive- and negative-energy
domains. The reflection coefficient ${\mathbb{R_+}}$
(\ref{refltr}), when extended to $\varepsilon<-m$ becomes greater
than unity, while the transmission coefficient $\mathbb{T_+}$
(\ref{refltr}) becomes negative. Facing such situation, known as
superradiation, we concluded that the spontaneous pair production
is impossible due to the Pauli ban: electrons, taken from the
Dirac sea, would have been increased in number after reflected off
the nucleus. This is, however, forbidden, since all the vacancies
in the Dirac sea are filled. We now renounce this point of view:
the reflection and transmission coefficients in the
negative-energy domain defined by (\ref{refltrneg}) and
(\ref{transnegative}) obey the inequalities $0\leq\mathbb{R_-}\leq
1,~~0\leq\mathbb{T_-}\leq 1$, the same as (\ref{refltr}),
(\ref{transpositive}) in the positive-energy domain. In other
words, the superradiation does not occur in our case of a spinor
particle, where it might forbid the spontaneous pair creation.

The following comparison with the known results on scattering of
Bose- and Fermi-particles off a black hole is in order. According
to \cite{chandra}, if the black hole is rotating and
correspondingly described by the Kerr metrics, the dilation
transformation, analogous to (\ref{standard}) that reduces the
corresponding propagation equation to a standard form of a
Schr$\ddot{\rm o}$dinger-like equation may, in a certain
kinematical domain, be two-fold-reversible both for
electromagnetic and gravitational waves, and the Dirac spin-$\frac
1{2}$ field. Simultaneously a singularity comes out from beyond
the horizon into the coordinate region where the differential
equation is defined. The general fact is that the current (or the
Wronsky determinant) changes its sign when crossing this
singularity for Bose-fields, but does conserve for Fermi-fields.
(We observed the same property of current conservation across the
singular point $r=-R$ in the differential equation
(\ref{standard}) in  Section \ref{Dilated}. Also this property holds
true for the singular barrier problem associated with the
second-order Schr$\ddot{\rm o}$dinger-like equation, to which the
Dirac equation in the Coulomb potential was reduced in \cite{zp}
and - after a coordinate-dilation transformation - in
\cite{shabad4}. Stress, that the singularity we are referring to
is not the original singularity of the initial equation, but the
one acquired in the course of its transformations.) Therefore, the
superradiation takes place where there is the current
discontinuity in the singularity point, $i.e.$ for Bose-particles.
\section{Absorption of electrons and production of
electron-positron pairs}\label{Absorption} We now turn to more
thorough consideration of expressions (\ref{transpositive}) and
(\ref{transnegative}), derived above, and of their physical
implementation.

The transmission coefficient (\ref{transpositive}), when
multiplied by an electron flux density, is the absorption
probability of electrons, incident on the point-like nucleus with
$Z>137$. Substituting (\ref{parameters}) in (\ref{transpositive})
we get the transmission/absorption coefficient as a function of
$\varepsilon$ and $Z$, $\mathbb{T_+}(\varepsilon,Z)$. It has  the
following asymptotic values
\bee\label{asymppos}\mathbb{T_+}(\infty,Z)=1-{\rm
e}^{-2\pi\gamma}\frac{\sinh\pi (\alpha Z-\gamma (Z))}{\sinh \pi
(\alpha Z+\gamma (Z))}<1,\nonumber\\ \mathbb{T_+}(m,Z)=1-{\rm
e}^{-4\pi\gamma(Z)}<1,\nonumber\\\quad
\mathbb{T_+}(\infty,Z)>\mathbb{T_+}(m,Z)\eend that
depend on the nucleus charge $Z$ (see Figure
1).\begin{figure}[htb]
  \begin{center}
   \includegraphics[bb = 0 0 405 210,
    scale=1]{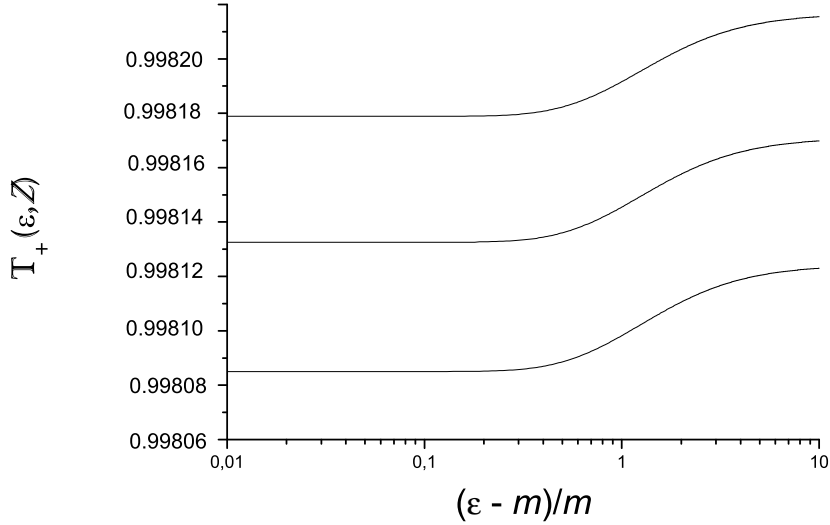}
\caption{Transmission/absorption coefficient
$\mathbb{T_+}(\varepsilon,Z)$ (\ref{transpositive}),
(\ref{parameters}) plotted in logarithmic scale against electron
kinetic energy $\frac{\varepsilon-m}{m}>1$ for three values of
nucleus charge, from bottom to top, $\gamma=0.498,~0.500,~0.502$
($\alpha Z=1.117,~1.118,~1.119)$. The transmissiom/absorption
coefficient is close to unity } \label{fig:1}
  \end{center}
\end{figure} Near the
threshold of absorption $\gamma=0~~(\alpha Z=1$) at the border of
sector IV these asymptotic values behave as \bee\label{asymppos2}
\left.\mathbb{T_+}(\infty,Z)\right|_{\alpha Z\rightarrow 1}\simeq
2\pi\sqrt{(\alpha Z)^2-1}~(1+\coth\pi)=4.008\pi\sqrt{(\alpha
Z)^2-1},\nonumber\\\left.\mathbb{T_+}(m,Z)\right|_{\alpha
Z\rightarrow 1}\simeq 4\pi\sqrt{(\alpha Z)^2-1}.\eend This means that near the very threshold the
absorption is very low. On the contrary, already for $\alpha
Z\simeq 1.12$ ($\gamma\simeq 0.5)$ the value of
$\mathbb{T_+}(\varepsilon,Z)$ exceeds 0.998 in the whole energy
range $m<\varepsilon<\infty$ (see Figures 1,~2).

The threshold behaviour of (\ref{transpositive}) for any energy
is\bee\label{threshold}\left.\mathbb{T_+}(\varepsilon,\gamma)\right|_{\alpha
Z\rightarrow 1}\simeq 2\pi\sqrt{(\alpha
Z)^2-1}~(1+\coth\frac{\pi\varepsilon}{\sqrt{\varepsilon^2-m^2}}),\quad
\varepsilon>m.\eend
\begin{figure}[htb]
  \begin{center}
   \includegraphics[bb = 0 0 405 210,
    scale=1]{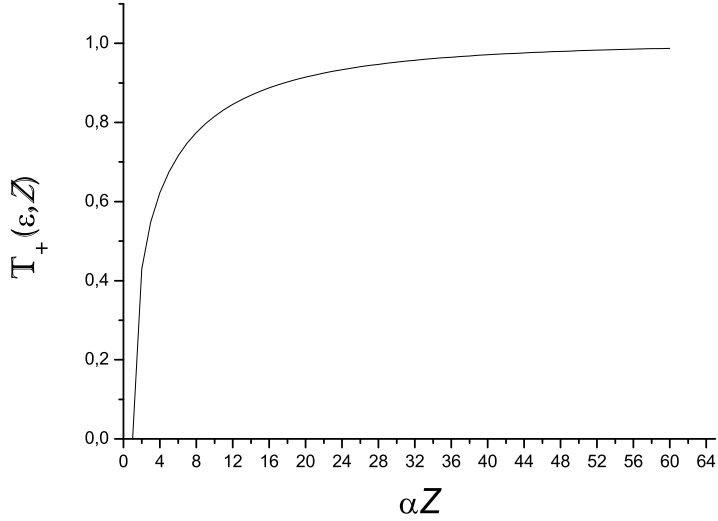}
\caption{Transmission/absorption coefficient
$\mathbb{T_+}(\varepsilon,Z)$ (\ref{transpositive}),
(\ref{parameters}) plotted against the nucleus charge for the
electron kinetic energy value $\varepsilon-m=11\cdot 10^{-6}m$.
The curves for other energy values are indistinguishable in the
scale of the figure from the plotted one: the difference with the
transmissiom/absorption coefficient for $\varepsilon-m=100~m$
makes about $10^{-4}$.} \label{fig:2}
  \end{center}
\end{figure}

\begin{figure}[htb]
  \begin{center}
   \includegraphics[bb = 0 0 405 210,
    scale=1]{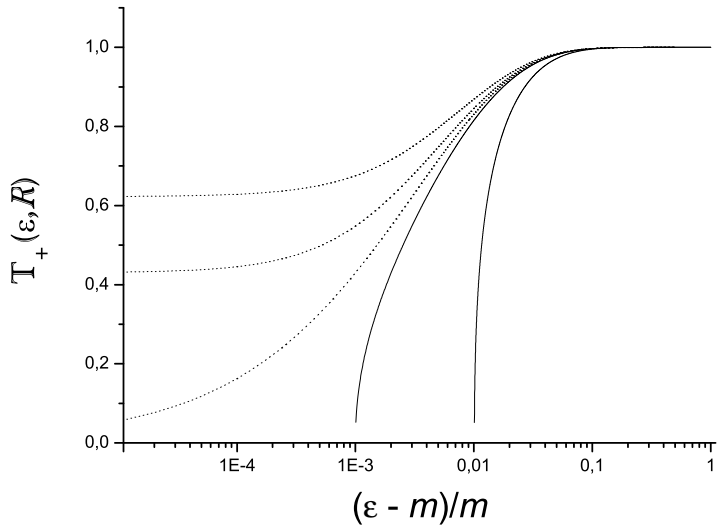}
\caption{Transmission/absorbtion coefficient
$\widetilde{\mathbb{T}}_+ (\varepsilon,R)$  (\ref{transpositive}),
(\ref{substitution}) plotted in logarithmic scale against electron
kinetic energy $\frac{\varepsilon-m}{m}>1$ for five values of the
ratio (\ref{ratio}), from left to right,
$Rm=1.003,~1.001,~1.000,~0.999,~0.990,$. The curves with $Rm<1$
(drawn solid) show the threshold behaviour (\ref{thr})}
\label{fig:3} \end{center}
\end{figure}

It is also of interest to consider the transmission/absorption
coefficient as a function of $\varepsilon$ and $R$, since the
Hilbert space is formed by solutions with fixed ratio
(\ref{ratio}), according to Section \ref{Bliss}. Call
$\widetilde{\mathbb{T}}_+(\varepsilon,R)$ the function, obtained
from (\ref{transpositive}) by the substitution ($R>0$, once
$\varepsilon\geq m$, $\alpha Z\geq
1$)\bee\label{substitution}\gamma=\sqrt{\varepsilon^2R^2-1},\quad
\zeta=\frac{\varepsilon^2R}{\sqrt{\varepsilon^2-m^2}}.\eend Now
the asymptotic
value\bee\label{value}\widetilde{\mathbb{T}}_+(\infty,R)=1 \eend
is universal for every $R$. A family of curves
$\widetilde{\mathbb{T}}_+(\varepsilon,R)$ is shown in Figure
\ref{fig:3}. 

If $R\geq m^{-1}$, the border of sector IV is at $\varepsilon=m$,
$\gamma\geq 0$. The values of
$\widetilde{\mathbb{T}}_+(\varepsilon,R)$ at this border
are\bee\label{border}\widetilde{\mathbb{T}}_+(m,R)=1-{\rm
e}^{-4\pi\sqrt{m^2R^2-1}},\; R\geq m^{-1}\eend and correspond to
crossings of the axis $\varepsilon -m$ by the family. (Note, that
the ordinate axis in Figure 3 corresponds to
$\varepsilon-m=10^{-5}m^{-1}$, and not to $\varepsilon -m=0$.)

If $R\leq m^{-1}$, the border of sector IV is at $\varepsilon\geq
m,~\gamma=0$. The curves in Figure \ref{fig:3} cross the abscissa
axis in the points $\varepsilon=\varepsilon_{\rm thr}\equiv
R^{-1}$, found from the equation $\gamma =0$. The threshold
behaviour of the transmission/absorption coefficient near these
points is \bee\label{thr}\left.\widetilde{\mathbb{T}}_+
(\varepsilon,R)\right|_{\varepsilon\rightarrow\varepsilon_{\rm
thr}=R^{-1}}\simeq 2\pi
(2R)^{1/2}\sqrt{\varepsilon-R^{-1}}(1+\coth\frac
\pi{\sqrt{1-m^2R^2}}),\nonumber\\ \varepsilon\geq R^{-1}>m.
 \eend The absorption takes place, for $R$ fixed, for electron
energies exceeding the threshold values $\varepsilon_{\rm
thr}=R^{-1}$.

In the negative-energy domain $\varepsilon\leq -m$ of sector IV,
$\alpha Z>1,~R<0$ the transmission coefficient
(\ref{transnegative}), when multiplied by the (degenerate)
Fermi-distribution of electrons in the Dirac sea, which is unity,
becomes the distribution of positrons, spontaneously produced from
the vacuum. \begin{figure}[htb]
  \begin{center}
   \includegraphics[bb = 0 0 405 210,
    scale=1]{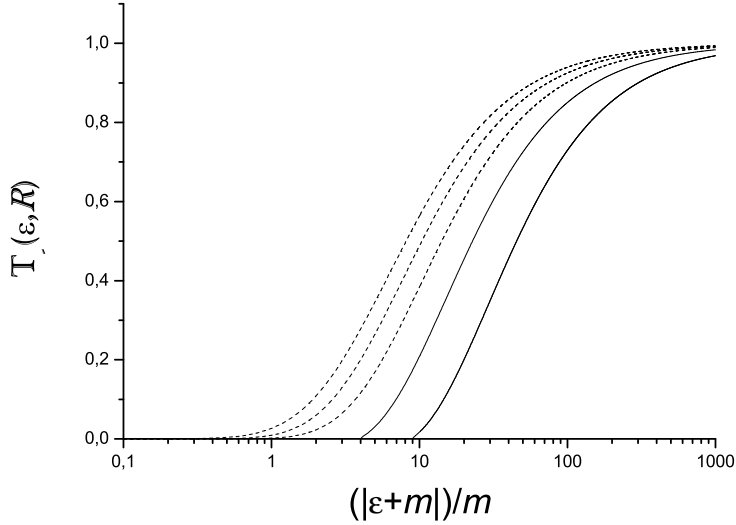}
\caption{Transmission coefficient/positron distribution
$\widetilde{\mathbb{T}}_- (\varepsilon,R)$  (\ref{transnegative}),
(\ref{substitution}) plotted in logarithmic scale against the
negative energy $\frac{\varepsilon+m}{m}<0$ for five negative
values of the ratio (\ref{ratio}), from  left to right, $Rm=-1,
~-2,~-3, ~-0.2,~-0.1$ The curves with $|R|m<1$ (drawn  solid) do
not gather in the origin and show the threshold behaviour
(\ref{thr2})} \label{fig:4}
  \end{center}
\end{figure}
These distributions as functions of energy,
$\widetilde{\mathbb{T}}_- (\varepsilon,R)$, are presented for the
fixed negative ratio $R$ in Figure 4, showing the transmission
coefficient (\ref{transnegative}), with (\ref{substitution})
substituted for $\gamma$ and $\zeta$ in it. The asymptotic value
for large $|\varepsilon
|\rightarrow\infty$\bee\label{totala}\left.\widetilde{\mathbb{T}}_-
(\varepsilon,R)\right|_{\varepsilon\rightarrow
-\infty}\simeq\rme^{2\pi(\zeta +\gamma)}\simeq\rme^{-\frac{\pi
m}{\varepsilon}(R m+\frac 1{R m})}\eend  is equal to unity for
every $R$: \bee\label{total} \widetilde{\mathbb{T}}_-
(-\infty,R)=1,\quad\; R<0.\eend This means that the total
probability of creating a positron with its energy less than
infinity is 1.

 Within the Dirac-sea picture, we are sticking to,
the mechanism, leading to the hole distributions like the ones
drawn in Figure 4, may be thought of as follows. According to
Section 2 (see the paragraph, preceding eq.(\ref{hamiltonian})),
the volumes of the inner and outer spaces unite in sector IV.
Consequently, the number of states is larger, than in sector II.
The states in the negative-energy part of sector II are all
occupied, whereas the newly added states are vacant. As a result,
electrons in the Dirac see rearrange: they tend to leave the
energy states with larger $|\varepsilon|$ and keep to the border
$\varepsilon=-m$. The larger $|\varepsilon|$, the more holes there
are.
%The resulting distribution of holes is thermal, but not black-body.

The surface $\widetilde{\mathbb{T}}_- (\varepsilon,R)$ is a
maximum for every energy at $R=-m^{-1}$. This fact is reflected in
Figure 4: the curve for $R=-m^{-1}$ occupies the extreme left
position. The pair creation process runs most efficiently, when
$R=-m^{-1}$, in other words, the distribution of produced
particles
%and, hence, the entropy
is maximum at this value of $R$.
The corresponding form of the distribution (\ref{transnegative})
is \bee\label{equilibrium}
\widetilde{\mathbb{T}}_-(\varepsilon,-m^{-1})=
\frac{1-\rme^{-4\pi\frac p{m}}}{\rme^{2\pi\frac
m{p}}-\rme^{-4\pi\frac p{m}}}.\eend Not close to the threshold
$p=0$, this is especially simple:
\bee\label{simple2}\widetilde{\mathbb{T}}_-(\varepsilon,-m^{-1})=
\rme^{-2\pi\frac m{p}}.\eend
%The "chemical potential" defined as
%the value of $\varepsilon$, where the second derivative of the
%distribution with respect to $\varepsilon$ changes its sign, is
%$\mu=-~3.14~m$.
The same as in the positive-energy domain
described above, the value $|R|=m^{-1}$ discriminates two
different situations. If $|R|>m^{-1}$, $R<0$, the border of sector
IV is $\varepsilon =-m,~\gamma\geq 0$, and the bordering value of
the distribution function (\ref{transnegative}) is
zero:\bee\label{zero1} \widetilde{\mathbb{T}}_-(-m,R)=0,\quad
|R|>m^{-1},\eend as is also seen in Figure 4: the curves with
$|R|>m^{-1}$ gather in the origin
$\widetilde{\mathbb{T}}_-=0,~\varepsilon +m=0$. If $|R|<m^{-1}$,
$R<0$, the border of sector IV is $\gamma =0, ~\varepsilon<-m$.
The threshold behaviour of the positron distribution is
\bee\label{thr2}\left.\widetilde{\mathbb{T}}_-(\varepsilon,R)
\right|_{\varepsilon\rightarrow\varepsilon_{\rm thr}=R^{-1}}\simeq
\pi (2)^{3/2}\sqrt{\varepsilon R-1}~\left(1+\coth\frac
\pi{\sqrt{1-m^2R^2}}\right),\nonumber\\ \varepsilon\leq R^{-1}<-m.
\eend This is zero in the threshold points $\varepsilon_{\rm
thr}=R^{-1}$, which are the intersections between the curves with
$R>-m^{-1}$ and the abscissa axis in Figure 4. By differentiating
(\ref{thr2}) over $\varepsilon$ we obtain that the differential
probabilities are singular, \bee\label{sing}\sim
(\varepsilon-R^{-1})^{-1/2},\eend near the threshold
$\varepsilon=\varepsilon_{\rm thr}$. The corresponding
singularities are seen as peaks against  the background of gentler
and wider maxima in Figure 5.
 \begin{figure}[htb]
  \begin{center}
   \includegraphics[bb = 0 0 405 210,
    scale=1]{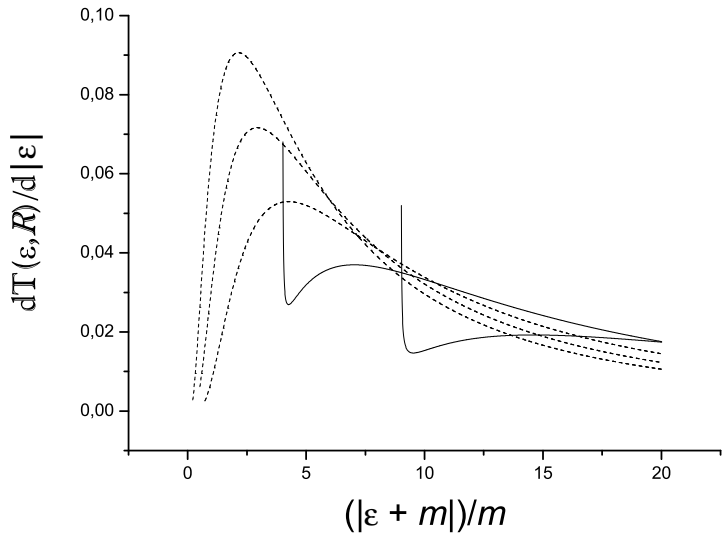}
\caption{Differential probability of positron creation -
derivative of the transmission coefficient (\ref{transnegative}),
(\ref{substitution}) ~$\rmd \widetilde{\mathbb{T}}_-
(\varepsilon,R)/\rmd |\varepsilon|$
 plotted  against the
negative energy $\frac{\varepsilon+m}{m}<1$ for five negative
values of the ratio (\ref{ratio}), from  left to right, $Rm=-1,
~-2,~-3,~-0.2,~-0.1$ The curves with $|R|m<1$ (drawn as solid)
show the resonant threshold behaviour (\ref{sing})} \label{fig:5}
  \end{center}
\end{figure}

We conclude that the absorption for $0<R<m^{-1},~\varepsilon>m$,
and positron production for $0>R>-m^{-1}, ~\varepsilon<-m$ are
resonant near the thresholds of the corresponding processes. Note,
that the condition $|R|<m^{-1}$ for the resonance to take place in
the pair production differential probability coincides with the
condition of the positivity of the norm discussed in Section 2.
\section{Discussion}\label{Discussion}
The existing theory of spontaneous $\rm e^+e^-$-pair creation by a
supercharged nucleus \cite{zp}, \cite{greiner} does not solve the
problem of singular interaction, but is satisfied with the
pragmatically sufficient
 view, that the realistic nucleus has a finite size, which provides the
 regularizing cut-off of the potential near $r=0$. The size of the nucleus
  core is taken to be about $10^{-12}$ cm, which is an order of magnitude
  less than the only characteristic size in the problem, the electron
  Compton wave-length $m^{-1}=3.9\cdot
10^{-11} cm$. Consequently, all the results
   in that theory - including the value of the critical charge itself -
   are cut-off dependent and do not survive its removal. Besides, as
   distinct from ours, those are based on numerical calculations. These
   circumstances forbid a direct comparison, because in our approach no
   cut-off is kept. Nevertheless we can point a certain
   correspondence between the spontaneous pair creation curves.

The theory of  \cite{zp}, \cite{greiner} points the critical
 value of the nucleus charge $Z$, the one for which the lowest electron
 level first sinks
  into the lower continuum $\varepsilon < - m$ and becomes a
  resonance, or a quasistationary state. The corresponding scattering
  amplitude acquires the Breit-Wigner shape. By considering a  nonstationary
problem, where the nucleus charge
 crosses its critical value in the course of time evolution, one establishes
  - with the use of the Fano theory \cite{greiner} - that the decay
  probability of the quasistationary state is determined by the same
   Breit-Wigner function. The quasistatinary state decays into a pair:
   the electron, which becomes a bound state localized near the centre,
   and the hole, which escapes to infinity and is interpreted as a free
   positron, subject to observation. (In our treatment, the created electron
    belongs to continuum of states, free near the center.) Thus,
    characteristic of the differential probabilities, obtained in this way,
     is a narrow Breit-Wigner resonance peak, its width depending on the cut-off.
      One may
      think, that the singular threshold behaviour seen in Figure 5 may be
      substituting for this resonance.

As for the electron absorption by the supercritical nucleus,
described in Section
 \ref{Absorption} and in \cite{shabad4}, this effect is unknown to the
conventional theory. According to Figures 1,2 the absorption runs very efficiently.
 It might be observed if the heavy ion collision process is subjected to irradiation
by a beam of electrons, or if the electrons produced in this process itself
via, $e.g.,$ the two-photon mechanism are absorbed.

There is another point, worth discussing. In our treatment above, the same
 as in the traditional treatment
of the pair-production process by a supercharged nucleus \cite{zp},
\cite{greiner}, the Dirac sea concept of the ground state and the hole
theory of positrons
were appealed to. It is known, however, that this concept cannot
 be done completely consistent, since the presence of a charge in the
 ground state remains not excluded within its scope. Nevertheless, this
 concept possesses a certain power of predictability and invokes useful
analogies loaned from the solid state physics and theory of phase
transitions. The interpretation of the produced positron
distribution as resulting from the adding of a manifold of vacant
states to the Dirac sea, proposed in Section \ref{Absorption}
above, may be referred to as an example of this sort.

In the meanwhile, the approach, developed in the present paper -
see particularly Section \ref{Dilated}, where the singular problem
is reduced to the barrier transmission/reflection on infinite axis
- suggests a natural context
 for applying the theory of unstable vacuum, used in the literature to
consider the Schwinger effect, $i.$$e.$ particle production by an
external electromagnetic field, \cite{nikishov}, \cite{gitman},
the Hawking radiation and Unruh's acceleration radiation
\cite{davies}. This theory is based on second quantization and
refers to nonequivalent Fock spaces, interrelated by the
Bogoliubov transformation, the Bogoliubov coefficients being just
the coefficients, that tangle
asymptotes with positive and negative  signs of momenta (in our
case also of pseudomomenta)
of the solutions to the equations of motion, prior to
the second quantizaton. Leaving the detailed elaboration of the
second-quantization
 programme of the singular problem to the forthcoming paper, we
 now restrict ourselves to formulating a clue point, important
 for description of the electron-positron pair production.

Let ~$a^{\pm}_{\rm out}$~ be operators, creating or annihilating
states, which possess the wave function that behaves as one of the
exponents (\ref{asympinfty}) at the spatial infinity
$r\rightarrow\infty$. Besides the Fock space, spanned by the
vectors created by repeatedly applying ~$a^{+}_{\rm out}$~ to the
out-vacuum ~~$|0_{\rm out}\rangle$, ~$ a^{-}_{\rm out}|0_{\rm out}
\rangle=0$,~ define another Fock space, produced by the action on
the true-vacuum ~$|0_{\rm true}\rangle$, ~$ a^{-}_{\rm
true}|0_{\rm true}\rangle=0$, ~of the operator ~$a^{+}_{\rm
true}$,~ which creates states, whose wave function is given by the
solution of the generalized eigenvalue problem, specified in
Section \ref{Bliss}. This eigenfuntion  behaves as a linear
combination of the two exponents (\ref{asympinfty}) at the spatial
infinity $r\rightarrow\infty$ (also a linear combination of the
two exponents (\ref{like}) or (\ref{likexi}) in the origin
$r\rightarrow 0$ or $\xi\rightarrow\pm\infty$) - \textit{cf}
reference \cite{shabad3}, where analogous eigenproblem was
explicitly solved for the Schr$\ddot{\rm o}$dinger equation with
singular potential. To determine the mean number of particles in
the true-vacuum state, suffices it to calculate the true-vacuum
expectation value of the particle-number operator ~~$ \langle
0_{\rm true} |a^{\dag-}_{\rm out}a^+_{\rm out}|0_{\rm
true}\rangle$. ~~This is expressed in terms of the coefficients in
the linear combination of the exponents mentioned and corresponds
to the count of particles by a remote observer. Simultaneously,
another observer, who is placed in the origin - or is infinitely
 remote in the dilated-coordinate $\xi$-space, - would measure
the mean number of particles ~~$\langle 0_{\rm
true}|a^{\dag-}_{\rm in}a^+_{\rm in}|0_{\rm true} \rangle$,~~where
~$a^{\pm}_{\rm in}$~ are operators, creating or annihilating
states with  the wave function that behaves as one of the
exponents (\ref{like}) or (\ref{likexi}) in the origin
$r\rightarrow 0$ or $\xi\rightarrow\pm\infty$. The gases of
particles, observed by the two observers, should be in mutual
balance. The idea of balance is formally introduced into the
theory, when we impose the non-Sturm boundary conditions
(\ref{period}) in the eigenvalue problem that interrelate the
values of the wave function in the points $r=0$ and $r=\infty$.
\ack The author is indebted to Yu.B.Khriplovich, who attracted the
author's attention to the fact that fermions usually do not
undergo superradiation, and to A.E.Lobanov and A.0.Barvinsky for a
valuable discussion. The work was supported  in part by the
Russian
 Foundation for Basic Research
 (project no 02-02-16944) and the President of Russia Programme
  for support of Leading Scientific Schools (LSS-1578.2003.2).
%\newpage

\end{document}